\documentclass[twoside]{article}
\usepackage{fleqn,espcrc2,epsf}

\usepackage{graphicx}
\usepackage[figuresright]{rotating}

\def\mxth{\mathsurround=0pt }
\def\xversim#1#2{\lower2.pt\vbox{\baselineskip0pt \lineskip-.5pt
  \ialign{$\mxth#1\hfil##\hfil$\crcr#2\crcr\sim\crcr}}}             
\def\gtrsim{\mathrel{\mathpalette\xversim >}}                                    \def\lesssim{\mathrel{\mathpalette\xversim <}}

\newcommand{\mweak}{M_{\rm{Weak}}}
\newcommand{\mplanck}{M_{\rm{Planck}}}
\newcommand{\hc}{\rm {H.c.}}

\newcommand{\AmS}{{\protect\the\textfont2
  A\kern-.1667em\lower.5ex\hbox{M}\kern-.125emS}}

\title{The scale of supersymmetry breaking as a free parameter
}

\author{N. Polonsky\address[MCSD]{Center for Theoretical Physics,
        Massachusetts Institute of Technology, \\ 
        Cambridge, Massachusetts 02139 , USA}
\thanks{
Talk given at the Thirty Years of Supersymmetry workshop 
held at the University of Minnesota, MN, October 2000.}
}
       
\begin{document}

\begin{abstract}

While supersymmetric extensions of the Standard Model
can be fully described in terms of explicitly broken 
global supersymmetry, this description is only effective.
Once related to spontaneous breaking
in a more fundamental theory, the effective parameters
translate to functions of two distinct scales,
the scale of spontaneous supersymmetry breaking 
and the scale of its mediation to the standard-model fields.
The scale dependence will be written explicitly and 
the full spectrum of supersymmetry breaking operators 
which emerges will be explored.
It will be shown that, contrary to common lore,
scale-dependent operators can play an important role
in determining the phenomenology.
For example, theories with low-energy supersymmetry breaking,
such as gauge mediation, may correspond to a scalar potential
which is quite different than in theories with  
high-energy supersymmetry breaking, such as gravity mediation.
As a concrete example, the Higgs mass prediction will be 
discussed in some detail and its upper bound will be shown to
be sensitive to the supersymmetry breaking scale.

\vspace{1pc}

\begin{center}
{\bf MIT-CTP-3085}
\end{center}

\end{abstract}

\maketitle

\section{INTRODUCTION}

If supersymmetry is realized in nature, and if it plays a role in
resolving the hierarchy problem associated with the quadratically divergent
quantum corrections to the Standard Model (SM) Higgs boson,
then it is explicitly broken (as dictated by experiment)
at the electroweak scale (as dictated by its solution to
the hierarchy problem).
The breaking is often assumed to be soft
so that the theory is at most logarithmically divergent and
no new hierarchy problem appears. 
Hence, the dynamics of the scalar fields $\phi$, for example, 
is described by a  potential of the form
\begin{eqnarray}
V &=& \left| \frac{\partial W}{\partial \Phi} \right|^{2}
+ \frac{g^{2}}{2}\left|\sum_{i}\phi_{i}T_{i}\phi_{i}^{\dagger}\right|^{2}
+ m^{2}\phi\phi^{\dagger} 
\nonumber  \\
&+&  
(B\phi^{2} + A\phi^{3} + A^{\prime}\phi^{2}\phi^{\dagger} + \hc). 
\label{Vssb}
\end{eqnarray}
In the first ($F$-)term, $W$ 
is the superpotential,\footnote{$\Phi = \phi +\theta\psi$ is 
a chiral superfield,
$\theta$ is the superspace coordinate, and $\psi$ is a fermion.}  
$W(\Phi)= \mu H_{1}H_{2} +$ Yukawa terms,
describing the interactions of the matter and Higgs fields,
with $H_{1}$ ($H_{2}$)
being the negative- (positive-)hypercharge Higgs doublet.
In the second ($D$-)term, $g$ is the gauge coupling,
and a summation over all gauge groups and generators $T$ is implied.
The $F-$ and $D-$terms dictate, in this description, all quartic interactions
which preserve (up to logarithmic corrections \cite{SO})
the supersymmetric structure.
All other terms break supersymmetry softly. 
Note that the soft supersymmetry breaking (SSB) parameters
are necessarily dimensionful (due to the softness requirement). 
Indeed, this is the minimal description of  supersymmetric
extensions of the SM, with the exception of the non-holomorphic
$A^{\prime}$-parameters which may not be soft in models containing
pure singlets (and which are often omitted).  
While the $F-$ and $D-$terms preserve supersymmetry
and depend on the gauge and Yukawa structure of the low-energy theory,
the SSB terms should depend
on the ultraviolet supersymmetry breaking vacuum expectation values (VEVs).
As such, they depend on, and encode, hidden ultraviolet physics.
Spontaneous supersymmetry breaking is most conveniently parameterized in terms
of a spurion field $X = \theta^{2}F$ with a non vanishing $F$-VEV,
which is an order parameter of (global) supersymmetry breaking. 
All SSB parameters
can be written as nonrenormalizable operators 
which couple the spurion $X$ to the SM superfields.
The operators are suppressed by the scale of
the mediation of supersymmetry breaking from the 
(hidden) sector parameterized by the spurion to the SM (observable) sector.
The coefficients of the various operators are dictated
by the nature of the interaction between the sectors and
by the loop-order at which it occurs. In the following, we will
explicitly write all operators which couple the two sectors.
This will allow us to consider a more general and
complete form of the supersymmetry breaking potential, which still
resolves the hierarchy problem. 
While we will also identify the operators 
which correspond to the SSB parameters,
our focus will be on those operators
which are not included in the minimal form of
the potential (\ref{Vssb}). 
This will be done in Sec.~\ref{sec:hard}.
Eventhough the SSB parameters appearing in (\ref{Vssb})
are functions of the relevant ultraviolet scales,
their magnitude is 
uniquely determined by the assumption that supersymmetry stabilizes
the weak scale against divergent quantum corrections,
$m \sim {\cal{O}}(\mweak)\sim {\cal{O}}(100)$ GeV etc.   
This places a constraint on the ratio
of the supersymmetry breaking scale $\sqrt{F}$ and the scale of
its mediation $M$ such that $F/M \simeq \mweak$,
and it implies that no information
can be extracted\footnote{Specific ultraviolet relations between
the various SSB parameters could
still be studied using the renormalization group formalism.}
regarding either scale from the SSB parameters.
(We will comment on the case of $A^{\prime}$ below.)
The generalization of (\ref{Vssb}) introduces dimensionless
hard supersymmetry breaking (HSB) parameters in the potential.
Their magnitude depends on 
the various scales in a way which both
provides useful information (unlike the SSB parameters)
and does not destabilize
the solution to the hierarchy problem. In Sec.~\ref{sec:higgs}
we will demonstrate this while calculating the Higgs mass,
which will provide a concrete example. We conclude in 
Sec.~\ref{sec:con}.

\section{CLASSIFICATION OF OPERATORS}
\label{sec:hard}
It is convenient to contain, without loss of generality,
all observable-hidden interactions in the non-holomorphic
Kahler potential
$K$, ${\cal{L}} 
= \int d^{2}\theta d^{2}\bar{\theta}$ (which is not protected by
nonrenormalization theorems).
We therefore turn to a general classification of $K$-operators,
originally presented in Ref.~\cite{N2}. (See also Ref.~\cite{martin}.)
We do not impose any global symmetries, which can obviously eliminate
some of the operators, and we keep all operators which 
survive the superspace integration. (Note that the spurion is
assumed to have only a $F$-VEV.)
The effective low-energy Kahler potential of a rigid 
$N = 1$ supersymmetry theory\footnote{This description
is useful also to $N=2$ supersymmetry under certain
assumptions \cite{N2}.} is given by
\begin{eqnarray}
K &=&  K_{0}(X,X^{\dagger}) + K_{0}(\Phi,\Phi^{\dagger}) 
\nonumber \\
&+& 
\frac{1}{M}K_{1}(X,X^{\dagger},\Phi,\Phi^{\dagger})
\nonumber \\
&+& \frac{1}{M^{2}}K_{2}(X,X^{\dagger},\Phi,\Phi^{\dagger})
\nonumber \\
&+& 
\frac{1}{M^{3}}K_{3}(X,X^{\dagger},\Phi,\Phi^{\dagger},D_{\alpha},W_{\alpha})
\nonumber \\
&+& 
\frac{1}{M^{4}}K_{4}(X,X^{\dagger},\Phi,\Phi^{\dagger},D_{\alpha},W_{\alpha})
+ \cdots 
\label{Kgeneral}
\end{eqnarray}
where, as before, $X$ is the spurion and 
$\Phi$ are the chiral 
superfields of the low-energy theory. 
$D_{\alpha}$ is the covariant derivative with respect 
to the superspace chiral coordinate $\theta_{\alpha}$, 
and $W_{\alpha}$ is the $N = 1$ gauge supermultiplet in its chiral 
representation, $W_{\alpha} \sim \lambda_{\alpha} + \theta_{\alpha} V$,
with $\lambda$ and $V$ here being the gaugino and vector boson, respectively.
Once a separation between supersymmetry breaking field $X$ and low-energy
$\Phi$ fields is imposed, there is no tree-level renormalizable
interaction between the two sets of fields, and their mixing
can arise only at the non-renormalizable level $K_{l \geq 1}$.
The superspace integration  ${\cal{L}}_{D} = 
\int d^{2}\theta d^{2}\bar{\theta}K$
reduces $K_{1}$ and $K_{2}$ to the usual SSB terms, as well as the 
superpotential $\mu$-parameter $W \sim \mu\Phi^{2}$, which were 
discussed in the previous section. It also contains Yukawa operators
$W \sim  y\Phi^{3}$ which can appear in the effective low-energy 
superpotential. These are summarized
in Tables~ \ref{table:o1} and~\ref{table:o1b}.
(We did not include linear terms that may appear
in the case of a singlet superfield.) Finally, The last 
term in Table~\ref{table:o1} contains correlated but unusual quartic 
and Yukawa couplings. 
They are soft as they involve at most logarithmic divergences.
Integration over $K_{3}$ produces non-standard soft terms, one of which 
was discussed in the previous section. 
These terms are soft unless the theory contains a pure singlet field,
in which case they can induce a quadratically divergent linear term.
They are summarized in Table~\ref{table:o2}.
The integration over $K_{3}$ 
also generates contributions to the (``standard'') 
$A-$ and gaugino-mass terms.
These terms could arise at lower orders in $\sqrt{F}/M$ from integration 
over holomorphic functions 
(and in the case of $A$, also from $K_{1}$). 
However, this  is equivalent to integration 
over $K$ if 
$\int d^{2}\bar{\theta}(X^{\dagger}/M^{2}) \simeq 1$.  
Note that in the presence of superpotential Yukawa couplings, a 
supersymmetry breaking
Higgsino mass term $\tilde{\mu}\widetilde{H}_{1}\widetilde{H}_{2}$
can be rotated to a combination of $\mu$ and ${A^{\prime}}$ terms,
and vice versa.  
Lastly, superspace integration over $K_{4}$ leads to dimensionless
hard operators. These are summarized in Table~\ref{table:o3},
and will occupy the remaining of this lecture.
Table~\ref{table:o3} also contain supersymmetry breaking gauge-Yukawa
interactions $\sim \phi^{*}\psi\lambda$. This is equivalent to 
the HSB kinetic term for the gauginos which was discussed recently
in Ref.~\cite{KK}. (HSB gaugino couplings are also generated radiatively
in the presence of SSB \cite{SO,HIK}.)
Higher orders in $(1/M)$ can be safely neglected
as supersymmetry and the superspace integration allow only
a finite expansion in $\sqrt{F_X}/M$, that is ${\cal{L}}
= f[F^{n}_X/M^{l}]$ with $n \leq 2$ and
$l$ is the index $K_{l}$ in expansion Eq.~(\ref{Kgeneral}). 
Hence, terms with $l > 4$ are suppressed by at least 
$(\langle X \rangle /M)^{l-4}$.
We will assume the limit $\langle X \rangle \ll M$
for the supersymmetry preserving VEV  $\langle X \rangle$, i.e., 
$X \sim \theta^{2}F_{X}$, 
so that all such operators can indeed be neglected
and the expansion is rendered finite.
\begin{table}
\caption{The soft supersymmetry breaking terms as operators
contained in $K_{1}$ and $K_{2}$. $\Phi = \phi + \theta\psi + \theta^{2}F$
is a low-energy superfield while $X$, $\langle F_{X} \rangle  \neq 0$, 
parameterizes supersymmetry breaking. 
$F^{\dagger} = \partial W/ \partial \Phi$.
The infrared operators are obtained by superspace integration
over the ultraviolet operators.
}
\label{table:o1}
\vspace*{0.3cm}
\begin{tabular}{cc}\hline
ultraviolet $K$ operator & infrared ${\cal{L}}_{D}$ operator  \\ \hline
&\\
$\frac{X}{M}\Phi\Phi^{\dagger} + \hc$ & $A \phi F^{\dagger}+ \hc$ \\
&\\
$\frac{XX^{\dagger}}{M^{2}}\Phi\Phi^{\dagger}$ + \hc& 
$\frac{m^{2}}{2}\phi\phi^{\dagger} + \hc $ \\
&\\
$\frac{XX^{\dagger}}{M^{2}}\Phi\Phi + \hc $ & $B\phi\phi + \hc$ \\
&\\
$\frac{X^{\dagger}}{M^{2}}\Phi^{2}\Phi^{\dagger}
+ \hc $ & $\kappa\phi^{\dagger}\phi F + \hc$ \\
 & $y\phi^{\dagger}\psi\psi + \hc$ \\

\end{tabular}
\end{table}
%

\begin{table}
\caption{
The effective renormalizable
$N = 1$ superpotential $W$ operators
contained in $K_{1}$ and $K_{2}$, ${\cal{L}} = \int d^{2}\theta W$. 
Symbols are defined in Table \ref{table:o1}.
The infrared operators are obtained by superspace integration
over the ultraviolet operators.
}
\label{table:o1b}
\vspace*{0.3cm}
\begin{tabular}{cc}\hline
ultraviolet $K$ operator & infrared $W$ operator  \\ \hline
&\\
$\frac{X^{\dagger}}{M}\Phi^{2}+ \hc$ & $\mu\Phi^{2}$ \\
&\\
$\frac{X^{\dagger}}{M^{2}}\Phi^{3}+ \hc$ & $y\Phi^{3}$ \\

\end{tabular}
\end{table}
%

\begin{table}
\caption{The non-standard or semi-hard  
supersymmetry breaking terms as operators
contained in $K_{3}$. 
$W^{\alpha}$ is the $N=1$ chiral
representation of the gauge supermultiplet and $\lambda$ is the respective
gaugino. $D_{\alpha}$ is the covariant derivative with respect to the 
(explicit) superspace coordinate $\theta_{\alpha}$.
All other symbols are as in Table \ref{table:o1}.
The infrared operators are obtained by superspace integration
over the ultraviolet operators.
}
\label{table:o2}
\vspace*{0.3cm}
\begin{tabular}{cc}\hline
ultraviolet $K$ operator & infrared ${\cal{L}}_{D}$ operator  \\ \hline
&\\
$\frac{XX^{\dagger}}{M^{3}}\Phi^{3} + \hc$ 
& $A\phi^{3} +\hc$ \\
&\\
$\frac{XX^{\dagger}}{M^{3}}\Phi^{2}\Phi^{\dagger} + \hc$ 
& ${A}^{\prime}\phi^{2}\phi^{\dagger} +\hc$ \\
&\\
$\frac{XX^{\dagger}}{M^{3}}D^{\alpha}\Phi D_{\alpha}\Phi + \hc $ & 
$\tilde{\mu}\psi\psi + \hc$\\ 
&\\
$\frac{XX^{\dagger}}{M^{3}}D^{\alpha}\Phi W_{\alpha} + \hc $ & 
$M^{\prime}_{\lambda}\psi\lambda + \hc$ \\ 
&\\
$\frac{XX^{\dagger}}{M^{3}}W^{\alpha} W_{\alpha} + \hc $ & 
$\frac{M_{\lambda}}{2}\lambda\lambda + \hc$ \\ 

\end{tabular}
\end{table}
%

\begin{table}
\caption{The dimensionless hard  
supersymmetry breaking terms as operators
contained in $K_{4}$. 
Symbols are defined as in Tables \ref{table:o1} and \ref{table:o2}.
The infrared operators are obtained by superspace integration
over the ultraviolet operators.
}
\label{table:o3}
\vspace*{0.3cm}
\begin{tabular}{cc}\hline
ultraviolet $K$ operator & infrared ${\cal{L}}_{D}$ operator  \\ \hline
&\\
$\frac{XX^{\dagger}}{M^{4}}\Phi D^{\alpha}\Phi D_{\alpha} \Phi + \hc$ 
& $y\phi\psi\psi  +\hc$ \\
&\\
$\frac{XX^{\dagger}}{M^{4}}\Phi^{\dagger}D^{\alpha}\Phi D_{\alpha} \Phi + \hc$ 
& $y\phi^{\dagger}\psi\psi  +\hc$ \\
&\\
$\frac{XX^{\dagger}}{M^{4}}\Phi D^{\alpha}\Phi W_{\alpha}  + \hc$ 
& $\bar{y}\phi\psi\lambda  +\hc$ \\
&\\
$\frac{XX^{\dagger}}{M^{4}}\Phi^{\dagger}D^{\alpha}\Phi W_{\alpha} + \hc$ 
& $\bar{y}\phi^{\dagger}\psi\lambda  +\hc$ \\
&\\
$\frac{XX^{\dagger}}{M^{4}}\Phi W^{\alpha} W_{\alpha} + \hc$ 
& $\bar{y}\phi\lambda\lambda  +\hc$ \\
&\\
$\frac{XX^{\dagger}}{M^{4}}\Phi^{\dagger}W^{\alpha} W_{\alpha} + \hc$ 
& $\bar{y}\phi^{\dagger}\lambda\lambda  +\hc$ \\
&\\
$\frac{XX^{\dagger}}{M^{4}}\Phi^{2}\Phi^{\dagger\, 2} + \hc$ 
& $\kappa(\phi\phi^{\dagger})^{2} +\hc$ \\
&\\
$\frac{XX^{\dagger}}{M^{4}}\Phi^{3}\Phi^{\dagger} + \hc$ 
& $\kappa\phi^{3}\phi^{\dagger} +\hc$ \\

\end{tabular}
\end{table}
It is interesting to identify two phenomenologically
interesting groups of terms in $K$,
(i) those terms which can break the chiral symmetries
and can generate Yukawa terms in the low-energy effective theory,
and (ii) new sources for quartic interactions.
The relevant chiral symmetry breaking 
terms in tables~\ref{table:o1} and~\ref{table:o2} 
can be identified with $A-$ and ${A^{\prime}}-$terms
which couple the matter sfermions
to the Higgs fields of electroweak symmetry breaking.
The chiral symmetry breaking originates in this case in the
scalar potential and propagates to the fermions at one loop \cite{RAD}.
More interestingly,
a generic Kahler potential is also found to contain 
tree-level chiral Yukawa couplings.
These include ${\cal{O}}(F_X/M^{2})$ supersymmetry conserving and soft
couplings and ${\cal{O}}(F^{2}_X/M^{4})$ hard chiral symmetry breaking 
couplings, leading to new avenues for fermion mass generation \cite{N2}.
Quartic coupling arise at  ${\cal{O}}(F_X/M^{2})$,
from supersymmetry conserving operators in  Table~\ref{table:o1}
(depending on $F_{\Phi}$), and at
${\cal{O}}(F^{2}_X/M^{4})$ from hard 
couplings in  Table~\ref{table:o3}. 
They can potentially alter the supersymmetry
conserving nature of the quartic potential in (\ref{Vssb}).
The relative importance of the HSB operators
relates to a more fundamental question:
What are the scales $\sqrt{F_X}$ and $M$? 
This will be addressed in Sec.~\ref{sec:higgs}.
However, before doing so we need to address a different
question regarding the potentially destabilizing properties
of the different HSB operators,
which relates to the nature
of the cut-off scale $\Lambda$. 
Indeed, one has to confirm that a given theory  is not destabilized
when the hard operators  are included,
an issue which is interestingly model independent.
In order to do so, consider the implications of the hardness of the operators
contained in $K_{4}$. Yukawa and quartic couplings
can destabilize the scalar potential by corrections $\Delta m^{2}$ 
to the mass terms of the order of
\begin{equation}
\Delta m^{2} \sim \left\{\begin{array}{c}
\frac{\kappa}{16\pi^{2}}\Lambda^{2} \sim 
\frac{1}{16\pi^{2}}\frac{F^{2}_{X}}{M^{4}}\Lambda^{2} \sim 
\frac{1}{16\pi^{2}c_{m}}m^{2}
\\
\\
\frac{y^{2}}{16\pi^{2}}\Lambda^{2} \sim 
\frac{1}{16\pi^{2}}\frac{F^{4}_{X}}{M^{8}}\Lambda^{2} \sim 
\frac{1}{16\pi^{2}c_{m}}m^{2}\frac{m^{2}}{M^{2}},
\end{array} \right.
\label{QD}
\end{equation}
where we identified $\Lambda \simeq M$ and $c_{m}$ is a dimensionless
coefficient omitted in Table~\ref{table:o1}, $m^{2}/2 = c_{m}F_{X}^{2}/M^{2}$.
The hard operators were substituted by the appropriate
powers of $F_{X}/M^{2}$.
Once $M$ is identified as the cut-off scale above which
the full supersymmetry is restored,
then these terms are harmless as the contributions are bound
from above by the tree-level  scalar  squared-mass parameters.
In particular, the softness assumption imposed 
on the supersymmetry breaking terms in (\ref{Vssb}) was
not necessary. 
(This observation extends to the case of non-standard soft operators
such as $A^{\prime} \sim F_{X}^{2}/M^{3}$ in the presence of a singlet). 
In fact, such hard divergent corrections are well known in  $N=1$ supergravity
with $\Lambda = M = \mplanck$, where they perturb
any given set of tree-level boundary conditions for the SSB 
parameters \cite{boundary}. 
Given the supersymmetry breaking scale in this case, $F \simeq \mweak\mplanck$,
the Yukawa (and quartic) operators listed below are 
proportional in these theories 
to $(\mweak/\mplanck)^{n}$, $n=1,\,2$, 
and are often omitted.
Nevertheless, such terms can shift any boundary conditions
for the SSB by ${\cal{O}}(100\%)$ \cite{boundary}
due to quadratically divergent corrections. 
We conclude that, in general, quartic couplings and chiral Yukawa couplings
appear once supersymmetry is broken, and if supersymmetry
is broken at low energy
then these couplings could be sizable yet harmless.
We will explore possible implications of the HSB quartic couplings
in the next section.
\section{THE HIGGS MASS vs. THE SCALE OF SUPERSYMMETRY BREAKING}
\label{sec:higgs}
In the previous section we have shown that,
in general, HSB quartic couplings $\kappa_{hard}$ arise
in the scalar potential
(from non-renormalizable operators in the Kahler potential, for example).
Assuming that the SSB
parameters are characterized by a parameter $m_{0} \sim 1\, {\rm TeV}$
then
\begin{equation}
\kappa_{hard} = 
\tilde{\kappa}_{h}\frac{F^{2}}{M^{4}} \simeq
\tilde{\kappa}_{h}(16\pi^{2})^{2n}\left(\frac{m_{0}}{M}\right)^{2},
\label{lambdahard}
\end{equation}
where $M$ is a dynamically determined scale parameterizing the communication
of supersymmetry breaking to the SM sector, which is distinct
from the supersymmetry breaking scale $\sqrt{F} \simeq
(4\pi)^{n}\sqrt{m_{0}M}$. 
The exponent $2n$ 
is the loop order at which the mediation of supersymmetry breaking 
to the (quadratic) scalar potential occurs. 
(Non-perturbative dynmaics
may lead to different relations that can be described instead 
by an effective value of $n$.)
The coupling $\tilde{\kappa}_{h}$ is an unknown dimensionless coupling
(for example, in the Kahler potential). As long as such quartic couplings
are not arbitrary but are related to the source of the SSB parameters
and are therefore described by (\ref{lambdahard}), then 
they do not destabilize the scalar potential
and do not introduce quadratic dependence on the ultraviolet cut-off
scale, which is identified with  $M$. 
Stability of the scalar potential only constrains
$\tilde{\kappa}_{h} \lesssim \min{\left((1/16\pi^{2})^{2n-1}, 1\right)}$
(though calculability and predictability are diminished).
The $F-$ and $D-$term-induced 
quartic potential in (\ref{Vssb}) gives
for the (pure $D$-induced) 
tree-level Higgs coupling, $V = \kappa h^{4}$,
\begin{equation}
\kappa = \frac{g\prime^{2} + g^{2}}{4}\cos^{2}2\beta,
\label{lambda}
\end{equation}
where
$\tan\beta \equiv\langle H_{2} \rangle / \langle H_{1} \rangle$,
and we work in the decoupling in which
one physical Higgs doublet $H$ is sufficiently heavy and
decouples from electroweak symmetry breaking 
while a second SM-like Higgs doublet is 
roughly given by $h \simeq H_{1}\cos\beta + H_{2}\sin\beta$,
and we conveniently defined
\begin{equation}
H_n=\left(
\begin{array}{c}
H_n^+\\(H_n^0+iA_n^0)/\sqrt{2}
\end{array}
\right).
\end{equation}
The HSB  coupling corrects this relation.
Given the strict tree-level upper bound that follows from (\ref{lambda}),
$m_{h^{0}}^{2} \leq M_{Z}^{2}\cos^{2}2\beta$, it is suggestive that
HSB may not be only encoded in, but also measured via, the Higgs mass.
We will explore this possibility, originally
pointed out and studied in Ref.~\cite{HM}, in this section.
In the case that supergravity interactions mediate
supersymmetry breaking from some ``hidden'' sector (where supersymmetry
is broken spontaneously) to the SM sector, one has $M = \mplanck$.
The corrections are therefore negligible whether the mediation
occurs at tree level ($n=0$) or loop level ($n \geq 1$) and can be 
ignored for most purposes. (For exceptions, see Refs.~\cite{martin,boundary}.)
In general, however, the scale of supersymmetry breaking is an 
arbitrary parameter and depends on the dynamics that mediate the SSB
parameters.
For example, it was shown recently that in the case of $N=2$ supersymmetry
one expects $M \sim 1\,{\rm TeV}$ \cite{N2}. 
Also, in models with extra large dimensions
the fundamental $\mplanck$ scale can be as low as  a few TeV,
leading again to $M \sim 1\,{\rm TeV}$.
(For example, see Ref.~\cite{pomarol}.) A ``TeV-type'' mediation scale
implies a similar supesymmetry breaking scale and provides an
unconventional possibility.
(For a discussion, see Ref.~\cite{N2}.) 
If indeed $M \sim 1\,{\rm TeV}$ then
$\kappa_{hard}$ given in (\ref{lambdahard}) is ${\cal{O}}(1)$
(assuming tree-level mediation (TLM) and
${\cal{O}}(1)$ couplings $\tilde\kappa_{h}$ in the Kahler potential).
The effects on the Higgs mass must be considered in this case.
Though one may argue that TLM models represent a theoretical extreme,
this is definitely a viable possibility.
A more familiar and surprising example is given by
the (low-energy) gauge mediation (GM) framework \cite{GM}. In GM, 
SM gauge loops communicate between
the SM fields and some messenger sectors, mediating the SSB potential.
The Higgs sector and the related operators, however, are poorly
understood in this framework \cite{MU} and therefore 
all allowed operators should be considered.
In its minimal incarnation (MGM) $2n = 2$, and $M \sim
16\pi^{2}m_{0} \sim 100\, {\rm TeV}$ parameterizes both the mediation 
and supersymmetry breaking scales. 
The constraint (\ref{QD}) corresponds
to $\kappa_{hard} \sim \tilde{\kappa}_{h} \lesssim 1/16\pi^{2}$
and the contribution of $\delta{\kappa_{hard}}$
to the Higgs mass could be comparable to the contribution of
the supersymmetric coupling (\ref{lambda}).
A particularly interesting case is that of non-perturbative messenger dynamics
(NPGM) in which case $n_{eff} = 1/2$,
$M \sim 4\pi m_{0} \sim 10\, {\rm TeV}$ \cite{4pi}, 
and the constraint 
on $\tilde{\kappa}_{h}$ is relaxed to 1. Now
$\kappa_{hard}\lesssim 1$ terms could
dominate the Higgs mass. The various frameworks are summarized in
Table \ref{table:frameworks}.
\begin{table*}[htb]
\caption{Frameworks for estimating $\kappa_{hard}$.
(Saturation of the lower bound on $M$ is assumed.)}
\label{table:frameworks}
\newcommand{\m}{\hphantom{$-$}}
\newcommand{\cc}[1]{\multicolumn{1}{c}{#1}}
\renewcommand{\tabcolsep}{2pc} 
\renewcommand{\arraystretch}{1.2} 
\begin{tabular}{@{}l|l|l|l|l}
\hline
&$n$&$\tilde{\kappa}_{h}$& $M$ & $\delta{\kappa}_{hard}$  \\ \hline
TLM &$0$&$\sim 1$& $\gtrsim m_{0}$  & $(m_{0}/M)^{2} \sim 1$ \\
NPGM & $1/2$ & $\sim 1$& $\gtrsim 4\pi m_{0}$  & $(4\pi m_{0}/M)^{2} \sim 1$ \\
MGM  &  $1$ & $\lesssim 1/16\pi^{2}$ & $\gtrsim 16\pi^{2}m_{0}$ &
$(4\pi m_{0}/M)^{2} \sim 1/16\pi^{2}$ \\ \hline
\end{tabular}\\[2pt]
\end{table*}
In order to address the $\beta$-dependence
of the HSB contributions (which is different from that of all other terms)
we recall the general two-Higgs-doublet model (2HDM)\cite{2hdm}.
The Higgs quartic potential can be 
written down as\footnote{
In the decoupling limit it simply reduces to the
SM with one ``light'' physical Higgs boson $h^{0}$,
$m_{h^0}^2=\kappa{v}^2$,
$\kappa=c_{\beta}^4\kappa_1+s_{\beta}^4\kappa_2+
2s_{\beta}^2c_{\beta}^2(\kappa_3+\kappa_4+\kappa_5)+
4c_{\beta}^3s_{\beta}\kappa_6+4c_{\beta}s_{\beta}^3\kappa_7$,
where $s_{\beta} \equiv \sin\beta$ and $c_{\beta} \equiv \cos\beta$,
and $v = \langle h \rangle$.
} 
\begin{eqnarray}
V_{\phi^{4}}&=&
\frac{1}{2}\kappa_1(H_1^{\dagger}H_1)^2
+\frac{1}{2}\kappa_2(H_2^{\dagger}H_2)^2 \nonumber \\
&+&\kappa_3(H_1^{\dagger}H_1)(H_2^{\dagger}H_2)
+\kappa_4(H_1^{\dagger}H_2)(H_2^{\dagger}H_1) \nonumber \\
&+&\left\{\frac{1}{2}\kappa_5(H_1^{\dagger}H_2)^2+[
\kappa_6(H_1^{\dagger}H_1) \right. \nonumber \\
&+& 
\left. \kappa_7(H_2^{\dagger}H_2)]H_1^{\dagger}H_2+{\rm h.c.}\right\}.
\label{Vphi4}
\end{eqnarray}
Allowing additional hard supersymmetry breaking quartic terms besides the 
usual gauge ($D$-)terms and loop contributions, 
$\kappa_{1\ldots{7}}$ can be written out 
explicitly as 
\begin{eqnarray}
\kappa_{1,2}&=&\frac{1}{2}(g\prime^2+g^2)+\kappa_{soft\,1,2}
+\kappa_{hard\,1,2},\\
\kappa_3&=&-\frac{1}{4}(g\prime^2 - g^2)+\kappa_{soft\, 3}
+\kappa_{hard\, 3}, \\
\kappa_4&=&-\frac{1}{2}g^2+\kappa_{soft\, 4}
+\kappa_{hard\, 4}, \\
\kappa_{5,6,7}&=& \kappa_{soft\, 5,6,7}
+\kappa_{hard\, 5,6,7},
\end{eqnarray}
where $g\prime$ and $g$ are the SM hypercharge and SU(2) gauge couplings, and 
$\kappa_{soft\,i}$ sums the loop effects
due to soft supersymmetry breaking effects $\sim \ln{m_{0}}$.
The effect of the
HSB contributions $\kappa_{hard\,i}$ 
is estimated next.
While Ref.~\cite{HM} explore the individual contribution
of each of the $\kappa_{hard\, i=1,\cdots , 7}$,
here we will assume, for simplicity, that 
$\kappa_{hard\, i} = \kappa_{hard}$ are all equal
and positive.
The squared Higgs mass $m_{h^0}^{2}$ reads in this case
\begin{equation}
m_{h^0}^2 =
M_Z^2\cos^{2}{2\beta}+\delta{m}^{2}_{loop}+(c_{\beta}+s_{\beta})^4v^2
\kappa_{hard},
\label{mhall}
\end{equation}
where $\delta{m}^{2}_{loop} \lesssim M_{Z}^{2}$ and $v = 174$ GeV
is the SM Higgs VEV.
(Note that no new particles or gauge interactions were introduced.)
Given the relation (\ref{mhall}), one can
evaluate the HSB contributions to the Higgs mass for
an arbitrary $M$ (and $n$). 
We define an effective scale
$M_{*} \equiv (M/(4\pi)^{2n}\sqrt{\tilde{\kappa}_{h}})
({\rm TeV} /m_{0})$.
The HSB contributions decouple  
for $M_{*} \gg  m_{0}$, and the results reduce to the MSSM limit
with only SSB (e.g., supergravity mediation).
However, for smaller values of $M_{*}$ the Higgs mass is 
dramatically enhanced.  For $M$=1 TeV and TLM or 
$M$=$4\pi$ TeV and NPGM, both of which correspond to $M_{*} \simeq 1$ TeV, 
the Higgs mass  could be as heavy as  475 GeV for 
$\tan\beta=1.6$ and 290 GeV for $\tan\beta=30$.
This is to be compared with 104 GeV and 132 GeV \cite{carena}, 
respectively, if HSB are either ignored or negligible.
(The SM-like Higgs boson $h^{0}$ may be as heavy as 180 GeV
in certain $U(1)^{\prime}$ models \cite{U1P} with only SSB.)
In the MGM case 
$\tilde\kappa_{h} \lesssim 1/16\pi^{2}$ so that
$M_{*}\sim 4\pi\, {\rm TeV}$
(unlike the NPGM where $M_{*} \sim 1$ TeV).
HSB effects are now more moderate but can increase the Higgs mass by 40 (10) 
GeV for $\tan\beta=1.6\ (30)$ (in comparison to the case with only SSB.)
Although the increase in the Higgs mass in this case
is not as large as in the TLM and NPGM
cases, it is of the same order of magnitude as, 
or larger than, the two-loop 
corrections due to SSB \cite{carena}, 
setting the uncertainty range on any such calculation.
%

%
\begin{figure}[t]
\begin{center}
\epsfxsize= 7.5 cm
\leavevmode
\epsfbox[40 220 520 600]{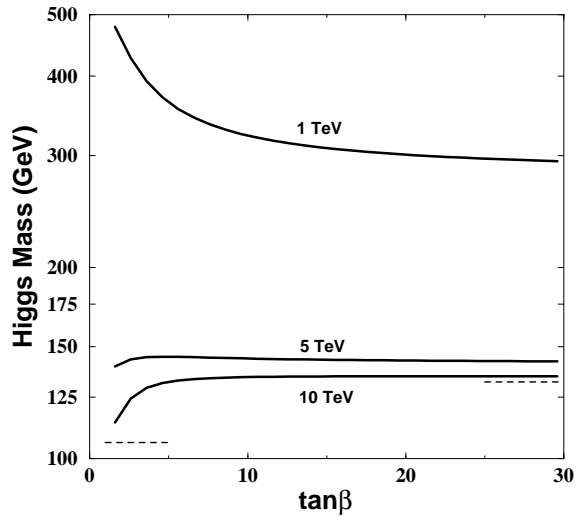}
\end{center}
\caption[f1]{The light Higgs boson mass (note the logarithmic scale)
is shown as a function of $\tan\beta$ for $M_{*}= 1,\,5,\,10$ TeV
(assuming equal HSB couplings).
The upper bound when considering only SSB ($M_{*} \rightarrow \infty$)
is indicated for comparison (dashed lines)
for $\tan\beta = 1.6$ (left) and $30$ (right).}
\label{fig:m1510}
\end{figure}

In Fig.~\ref{fig:m1510}, $m_{h^{0}}$ dependence on $\tan\beta$ for fixed 
values of $M_{*}$
is shown.  The $\tan\beta$ dependence is from
the tree-level mass and from the HSB corrections, 
while the loop corrections to $m_{h^{0}}^2$ are fixed, for simplicity,
at $9200\, {\rm GeV}^2$ \cite{carena}.
The upper curve effectively corresponds to $\kappa_{hard} \simeq 1$.
The HSB contribution dominates the Higgs mass and 
$m_{h^{0}}$ decreases with increasing $\tan\beta$.  As indicated above,
$m_{h^{0}}$ could be in the range of $300\ -\ 500$ GeV,
dramatically departing from  calculations
which ignored HSB terms.
The lower two curves illustrate the range\footnote{
Given the many uncertainties, e.g., the messenger quantum numbers
and multiplicity and $\sqrt{F}/M$ \cite{GM}, we identify the MGM with
a $M_{*}$-range which corresponds to a factor of two
uncertainty in the hard coupling.
} 
of the corrections in the MGM, 
where the tree-level 
and the HSB contributions compete.  
The $\cos2\beta$ dependence of the tree-level term
dominates the $\beta$-dependence of these two curves. 
Clearly, 
the Higgs mass could discriminate 
between the MGM and NPGM and help to better
understand the origin of the supersymmetry breaking. 
Following the Higgs boson discovery, 
it should be possible to extract information on the mediation scale $M$.
In fact, some limits can already be extracted.
Consider the upper bound on the Higgs mass derived from a fit to electroweak 
precision data: $m_h^{0}\ <\ 215$ GeV at 95$\%$ confidence level \cite{pdg}.
(Such fits are valid in the decoupling limit discussed here.) 
A lower bound on the scale $M$ in MGM could be obtained from 
\begin{eqnarray}
m_Z^2\cos^22\beta  + \delta{m}^2_{loop} 
(c_{\beta} + s_{\beta})^4v^2\left(\frac{4\pi{m}_0}{M}\right)^2 &&  \nonumber \\
\leq  (215\ {\rm GeV})^2 &&,
\label{limit}
\end{eqnarray}
assuming equal $\kappa_{hard}$'s.  
For $\beta=1.6$, it gives 
$M\geq$ 31 TeV while for $\tan\beta=30$ the lower bound is $M\geq$ 19 TeV.
Once $m_{h^{0}}$ is measured, 
more stringent bounds on $M$ could be set. 
In conclusion,
this section  illustrates that the scale of the mediation of
supersymmetry breaking explicitly appears in the prediction
of the Higgs mass (and with a distinct $\beta$-dependence).
In turn, it could lead in certain cases
to a much heavier Higgs boson than usually anticipated
in supersymmetric theories.
It could also distinguish models, e.g., supergravity mediation from 
other low-energy mediation and 
weakly from strongly interacting messenger sectors.
Given our ignorance of the (Kahler potential and) HSB terms,
such effects can serve for setting the uncertainty on any
Higgs mass calculations and can be used to
qualitatively constrain the scale of mediation of supersymmetry breaking
from the hidden to the SM sector.
\section{SUMMARY}
\label{sec:con}
The most general treatment of the mediation of supersymmetry
breaking requires  the parameterization of the various SSB and HSB
parameters in terms of the scale of supersymmetry breaking
and of the scale of its mediation to the low-energy fields.
This introduces in the infrared potential
supersymmetry breaking parameters whose
magnitude is not fixed by the weak scale, yet they do not destabilize
the weak scale. While in some cases the parameters are very small,
their effects need not be negligible. For example, in supergravity
mediation they could lead to sizeable correction to the SSB parameters.
In theories of low-energy supersymmetry breaking, 
such as gauge mediation,
the effects could be
even more dramatic, as illustrated in our discussion of the Higgs mass.
Furthermore, the effects of, e.g., quartic couplings $\sim F^{2}/M^{4}$
may allow one to indirectly measure the scale $M$, as well as to determine
the dynamics responsible for supersymmetry breaking
(as in the case of strongly $vs.$ weakly interacting messenger sectors).
Other interesting possibilities not explored here include
chiral symmetry breaking and fermion mass generations \cite{N2};
theories of extended supersymmetry \cite{N2} and
their accommodation of experimental constraints \cite{EX};
non-holomorphic trilinear interactions and their effect
on the stability of the vacuum \cite{RAD}; supergravity boundary
conditions \cite{boundary}; Higgsino mass generation \cite{MU};
theories with singlets \cite{SING};
and stabilization of flat potentials \cite{martin}.
The relevance of various issues depends on
the scale of supersymmetry breaking, which for the time being remains
a free parameter. Treating it as such can illuminate some of
the mysteries of the superworld as well as help in identifying
various phenomena.
It is pleasure to thank Shufang Su for her collaboration
and contribution to the works reviewed here.
This work is supported by 
the US Department of Energy under cooperative research 
agreement No.~DF--FC02--94ER40818. 

\end{document}